\documentclass[twocolumn, showpacs, prl]{revtex4}

\usepackage{graphicx}

\usepackage{dcolumn}

\usepackage{bm}

\begin{document}

\title{Realization of an Interacting Two-Valley AlAs Bilayer System}

\date{\today}

\author{K. Vakili}

\author{Y. P. Shkolnikov}

\author{E. Tutuc}

\author{E. P. De Poortere}

\author{M. Shayegan}

\affiliation{Department of Electrical Engineering, Princeton
University, Princeton, NJ 08544}

\begin{abstract}

By using different widths for two AlAs quantum wells comprising a
bilayer system, we force the X-point conduction-band electrons in
the two layers to occupy valleys with different Fermi contours,
electron effective masses, and g-factors. Since the occupied
valleys are at different X-points of the Brillouin zone, the
interlayer tunneling is negligibly small despite the close
electron layer spacing. We demonstrate the realization of this
system via magneto-transport measurements and the observation of a
phase-coherent, bilayer $\nu$=1 quantum Hall state flanked by a
reentrant insulating phase.

\end{abstract}


\maketitle

Two-dimensional (2D) electron systems subjected to large
perpendicular magnetic fields exhibit a wealth of phenomena, such
as the fractional quantum Hall effect, that are associated with
electron-electron interactions. When two 2D electron systems are
brought in close proximity, the additional, interlayer interaction
can lead to new many-body states that have no analogue in the
single-layer case. Examples include quantum Hall states (QHSs) at
even-denominator fillings $\nu$=1/2 and 3/2 \cite{suen92,
eisenstein92} and a special, bilayer $\nu$=1 QHS with interlayer
phase coherence \cite{murphy94} ($\nu$ is the Landau level filling
factor of the bilayer system). Such states form when the
interlayer distance is on the order of or smaller than the
magnetic length. It is also often desirable to have as little
interlayer tunneling as possible so that, e.g., independent
contacts can be made to the two layers \cite{spielman00};
moreover, negligible tunneling makes the theoretical treatment of
the phenomena in these systems easier.

We report here the fabrication of a novel bilayer system comprised
of two AlAs quantum wells (QWs) with different widths, wherein the
electrons in the two layers occupy different conduction-band
valleys. The key to the fabrication of our sample is the
following.  Bulk AlAs has an indirect band-gap with the conduction
band minima at the X-points of the Brillouin zone. The constant
energy ellipsoids (or valleys) formed at these minima are
anisotropic with two characteristic effective masses (measured in
units of the free electron mass): {\it m}$_{t}$=0.2 for the two
transverse directions, and {\it m}$_{l}$=1 for the longitudinal
direction. This is somewhat similar to Si, except that in Si there
are six ellipsoids centered around six equivalent points along the
$\Delta$-lines of the Brillouin zone, while in AlAs we have three
(six half-) ellipsoids at the X-points. When electrons are
confined along the growth ({\it z}) direction in an AlAs QW, one
might expect that only the out-of-plane (X$_{Z}$) valley would be
occupied because the larger electron mass along the confinement
direction should lower the energy of this valley. This is indeed
the case in Si MOSFETs and QWs. However, in AlAs QWs grown on GaAs
substrates, the strain induced by the lattice mismatch between
AlAs and GaAs causes the in-plane valleys to be occupied, unless
the QW is narrower than a threshold value of approximately 55
$\AA$ \cite{vankesteren89,lay93, yamada94, depoortere00}. By
growing a modulation-doped, double QW sample with well widths on
either side of this threshold, we force the electrons in each AlAs
QW to occupy differently oriented valleys. Moreover, in our
sample, anisotropic strain in the plane lifts the degeneracy
between the in-plane valleys so that, in the wider QW, only one of
the valleys (X$_{X}$) is occupied \cite{lay93,yamada94,
depoortere00}.

Because the occupied valleys are at different points of the
Brillouin zone, the interlayer tunneling is strongly suppressed
even though the layers are very closely spaced. Another novel
aspect of this system is that the Fermi contour shapes, effective
masses, and g-factors in the two layers are different. We
demonstrate the realization of this system via magneto-transport
measurements.  We also report the observation of a phase coherent
QHS at filling factor $\nu$=1, surrounded by a reentrant
insulating phase. In our sample, the $\nu$=1\ QHS is insensitive
to the application of an in-plane magnetic field, evincing that
tunneling is indeed negligible despite the very small interlayer
separation.

We studied a Si-modulation doped AlAs bilayer grown by molecular
beam epitaxy on a GaAs (100) substrate.  The structure consists of
a 100 $\AA$ wide QW in the front and a 45 $\AA$ QW in the back
(substrate side), surrounded by Al$_{0.4}$Ga$_{0.6}$As barriers
and separated by a 28 $\AA$ GaAs barrier, giving a QW
center-to-center separation of approximately 100 $\AA$.  The
sample was lithographically patterned in a Hall bar configuration,
and ohmic contacts to both layers were made by depositing a AuGeNi
layer and alloying in a reducing atmosphere. Metallic front and
back gates were added to allow for adjustment of the total charge
density in the bilayer ({\it n}$_{tot}$) and the densities in the
narrow and wide wells ({\it n}$_{N}$ and {\it n}$_{W}$)
individually. We determined {\it n}$_{tot}$, {\it n}$_{N}$, and
{\it n}$_{W}$ from a combination of Shubnikov-de Haas (SdH) and
Hall measurements. Typical values of {\it n}$_{tot}$ in our
experiment were in the range of 2 to 5 x 10$^{11}$ cm$^{-2}$ with
a mobility of $\sim$ 2.2 m$^{2}$/Vs. We made measurements in
$^{3}$He and dilution refrigerators with base temperatures of 300
mK and 40 mK, respectively. In both refrigerators, the sample was
mounted on a tilting stage so that the angle, $\theta$, between
the normal to the sample and the magnetic field could be varied
from 0$^{o}$ to 90$^{o}$ $\textit{in situ}$. We used standard
low-frequency lock-in techniques for transport measurements. Prior
to measurements, the sample was illuminated with a red LED while a
positive bias was applied to the gates in order to establish ohmic
contact with the wells as detailed in \cite{depoortere03}.

\begin{figure}
    \centering
    \includegraphics[scale=0.375]{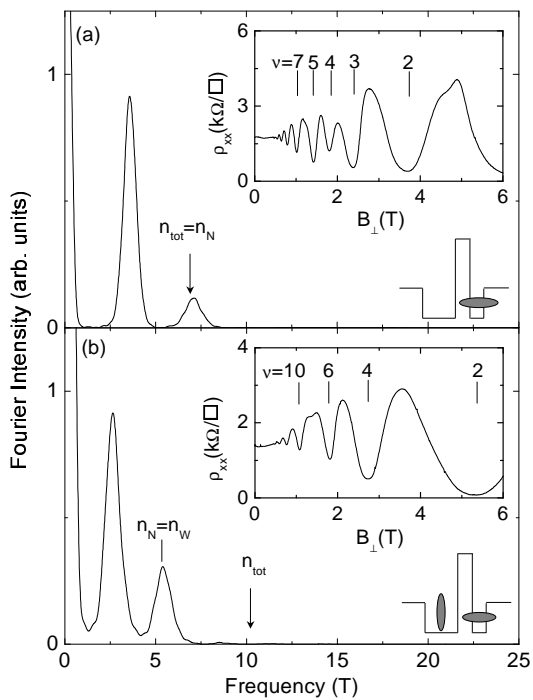}
    \caption{Fourier transform spectra of the Shubnikov-de Haas oscillations (shown in the insets)
    for different values of front and back gate biases.  The arrows indicate the expected peak positions,
    corresponding to the total bilayer density {\it n}$_{tot}$ (1.7x10$^{11}$ cm$^{-2}$ and 2.5x10$^{11}$ cm$^{-2}$ in the top and bottom traces, respectively), as determined from the Hall resistance; the bilayer total filling factor $\nu$ is also marked by vertical lines in the insets.
    As schematically shown by the lower insets, in (a) the electrons are all in the narrow well and occupy the X$_{Z}$ valley, while in (b), the electrons equally occupy the X$_{Z}$ valley of the narrow well and the X$_{X}$ valley of the wide well.}
\end{figure}

We first demonstrate how we change and monitor the densities in
each QW.  In Fig.\ 1 we show the Fourier transform (FT) spectra
(main figures) of SdH oscillations (upper insets) at different
values of front and back gate biases ({\it V}$_{FG}$ and {\it
V}$_{BG}$), corresponding to different {\it n}$_{W}$ and {\it
n}$_{N}$. The X-point conduction band energies and the orientation
of the occupied valleys in each layer are shown schematically in
the lower insets of each panel.  For each pair of {\it V}$_{FG}$
and {\it V}$_{BG}$, we also measure {\it n}$_{tot}$ from the slope
of the Hall resistance at low magnetic fields. To make the
connection to the frequency of the SdH oscillations, we divide the
measured {\it n}$_{tot}$ by the Landau level degeneracy, {\it
e/h}, and show the resulting frequency by a vertical arrow marked
as {\it n}$_{tot}$. Figure 1(a) corresponds to the simplest case
where all the electrons reside in the narrow QW. We observe strong
SdH oscillations, with $\rho$$_{xx}$ minima at every integer
$\nu$$\leq$5. Surprisingly, the minima at even-integer fillings
are weaker than those at odd-integer fillings, and disappear at
the lowest fields. This is consistent with our measurements of 2D
electrons occupying the X$_{Z}$ valley in single, narrow AlAs QWs,
and comes about because of the ratio ($\sim$ 0.7) of the Zeeman
and cyclotron energies. The FT spectrum of the oscillations
exhibits two peaks as commonly observed in a single-layer 2D
electron system \cite{tutuc02}: a higher frequency peak stems from
the spin-resolved Landau levels (at sufficiently large fields) and
its position coincides with {\it n}$_{tot}$, while a peak at half
this frequency originates from the spin-unresolved Landau levels
(at low fields). In Fig.\ 1(b), we have increased {\it n}$_{tot}$
and shifted the balance of electrons between the two layers so
that now {\it n}$_{N}$ = {\it n}$_{W}$. The SdH trace and its FT
spectrum are characteristic of a balanced bilayer electron system
with negligible interlayer tunneling, namely, two 2D layers in
parallel: $\rho$$_{xx}$ minima are observed only at even integer
fillings (for $\nu$ $>$ 1), and the FT spectrum does not show a
peak at {\it n}$_{tot}$ but rather at a value corresponding to
{\it n}$_{tot}$/2. We also have SdH and Hall data for other
configurations where the electrons are distributed unevenly
between the two layers. From such data, we can determine {\it
n}$_{tot}$, {\it n}$_{N}$, and {\it n}$_{W}$. However, we are not
able to shift all the electrons into the wide QW because of
experimental limitations on the maximum gate biases we could apply
to the sample.

Next, we measured the carrier effective mass ({\it m}*), from the
temperature dependence of the amplitude of the SdH oscillations,
for the case where all the electrons reside in the narrow QW. At a
density of {\it n}$_{N}$ = 1.8x10$^{11}$ cm$^{-2}$, we measured a
mass of 0.44$\pm$0.03, much larger than {\it m}*=0.2 expected for
in-plane motion if the electrons occupy the X$_{Z}$ valley. This
surprising result, however, is consistent with our results in
single, narrow AlAs QW samples that have shown a dependence of
{\it m}* on the density: in a single, 45 $\AA$ wide AlAs QW, we
have measured an {\it m}* that changes from 0.23 at a high density
of 7.2x10$^{11}$ cm$^{-2}$ to 0.44 at a low density of
1.8x10$^{11}$ cm$^{-2}$. A similar {\it m}* enhancement at low
densities was also reported in Si-MOSFETs \cite{shashkin02} where
the enhancement is attributed to electron-electron interaction. We
could not measure {\it m}* in our wider AlAs QW since we were not
able to put all the electrons in this well only. However, for an
in-plane valley, electron-electron interactions can enhance the
effective electron mass beyond the expected bare value of  {\it
m}*=({\it m}$_{l}${\it m}$_{t}$)$^{1/2}$=0.45, and our
measurements of mass in single wide AlAs QWs are consistently
higher than this bare value. It is therefore likely that, in our
balanced bilayer system [Fig.\ 1(b)], {\it m}* in the wider QW is
larger than {\it m}* in the narrow QW.

\begin{figure}
    \centering
    \includegraphics[scale=0.43]{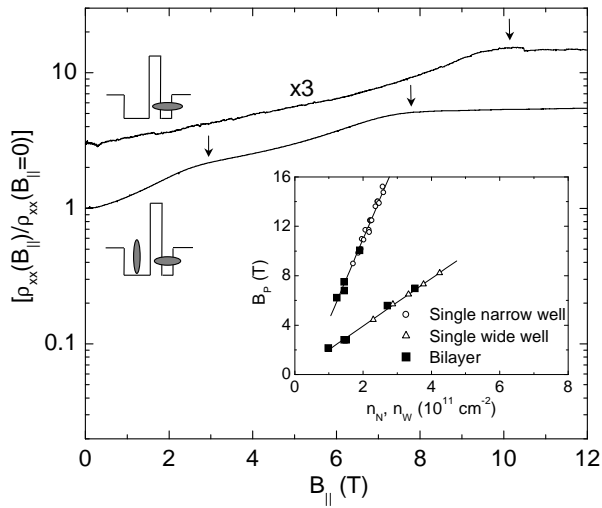}
    \caption{Parallel field magneto-resistance for different charge distributions; top trace: {\it n}$_{N}$ = 1.9x10$^{11}$ cm$^{-2}$,
    {\it n}$_{W}$ = 0; bottom trace: {\it n}$_{N}$ = {\it n}$_{W}$ = 1.5x10$^{11}$ cm$^{-2}$. The arrows mark the fields, {\it B}$_{P}$, above which the layer(s) become
    fully spin-polarized.  The inset shows {\it B}$_{P}$ versus layer density.}
\end{figure}

Our most convincing evidence establishing that the electrons in
our sample occupy different valleys in the two layers is provided
by the response of the system to a magnetic field, {\it B}$_{||}$,
applied parallel to the sample plane.  In Fig.\ 2 we show
magneto-resistance traces, as a function of {\it B}$_{||}$, for
two distributions of charge, similar to Fig.\ 1: one in which all
the electrons are in the narrow QW, and one where the charge
density is distributed evenly between the two QWs. In both cases,
there are pronounced kinks in the magneto-resistance traces,
indicated by the vertical arrows in Fig.\ 2. Previous
investigations in single-layer 2D systems, both theoretical and
experimental \cite{dolgopolov00,tutuc01}, have identified the
position of this kink ({\it B}$_{P}$) as the onset of complete
spin polarization of the carriers, that is the field beyond which
the Zeeman energy exceeds the Fermi energy of the 2D system.  In a
simple picture, where a linear spin polarization with {\it
B}$_{||}$ is assumed, the field at which the full polarization
occurs is determined by the product {\it g}*{\it m}* ({\it g}* is
the effective g-factor). In the inset to Fig.\ 2, we have plotted
the measured values of {\it B}$_{P}$ as a function of {\it
n}$_{N}$ and {\it n}$_{W}$, determined from the Hall and SdH data
for different pairs of {\it V}$_{FG}$ and {\it V}$_{BG}$. In the
same figure, we have also included {\it B}$_{P}$ measured in
$\textit{single}$ narrow and wide AlAs QWs \cite{depoortere02}.
The data fall on two distinct curves, approximately straight
lines, but with very different slopes. We will discuss the
quantitative details and implications of this data in a future
communication.  It is clear, however, that {\it B}$_{P}$ is nearly
a factor of two larger for electrons occupying the X$_{Z}$ valley
(narrow QW) than for those occupying the X$_{X}$ (wide QW) valley,
implying that {\it g}*{\it m}* is twice as small in the former
case \cite{finitelayer}. This conclusion is corroborated by our
measurements in tilted magnetic fields where we observe crossings
of the Landau levels at particular tilt angles, $\theta$, whose
values are a measure of the product {\it g}*{\it m}* of the 2D
system. In the narrow AlAs QW of the present sample, or in our
single, narrow AlAs QWs, we observe crossings at values of
$\theta$ that are very different from the values at which the
Landau levels in our wide AlAs QWs cross. We emphasize that, in
addition to the different mass and g-factor, there is a basic
difference in the shapes of the (in-plane) Fermi contours for the
two layers: circular for the X$_{Z}$ valley in the narrow QW and
elliptical for the X$_{X}$ valley in the wide QW.

Next, we present evidence that interlayer tunneling is small in
our bilayer and that the system exhibits phenomena arising from
strong interlayer interaction. In Fig.\ 3(a) we show $\rho_{xx}$
versus perpendicular magnetic field traces for a balanced state of
the bilayer at three different temperatures. Up to B $\sim$ 6 T,
QHSs are observed only at even $\nu$, consistent with two 2D
parallel layers with little or no interlayer tunneling. At
$\nu$=1, however, a strong $\rho_{xx}$ minimum is observed, which
we associate with the phase-coherent, bilayer QHS. Such a QHS has
been reported in GaAs bilayer electron \cite{murphy94, spielman00}
and hole \cite{hyndman96, tutuc03} systems; it is stabilized by
strong interlayer interaction, exists even in the limit of
vanishing interlayer tunneling, and occurs when {\it d/l}$_{B}$ is
less than about 1.6, where {\it d} is the distance between the
electron layers and {\it l}$_{B}$ is the magnetic length.  For the
trace in Fig.\ 3(a), {\it d/l}$_B$=1.27, consistent with these
previous reports. From the temperature dependence of $\rho_{xx}$
at $\nu$=1, we obtain an energy gap of approximately 1 K for this
state.

In Fig.\ 3(a), we also see the emergence of a reentrant insulating
phase (IP) around the $\nu$=1 minimum, extending from
approximately $\nu$=1.29 to 1.04 and beyond $\nu$=0.97. Previous
results have suggested that this phase may represent a pinned,
bilayer, Wigner crystal state, where interlayer interaction
stabilizes the state at a filling that is higher than what would
be expected based on single-layer data \cite{manoharan96,tutuc03}.
We note that the presence of the IP near the $\nu$=1 minimum in
the present bilayer system is consistent with previous
experimental results.  In particular, interacting GaAs bilayer
holes, whose effective mass is comparable to the mass of AlAs
electrons, show an IP that is also reentrant around $\nu$=1
\cite{tutuc03}.

\begin{figure}
    \centering
    \includegraphics[scale=0.375]{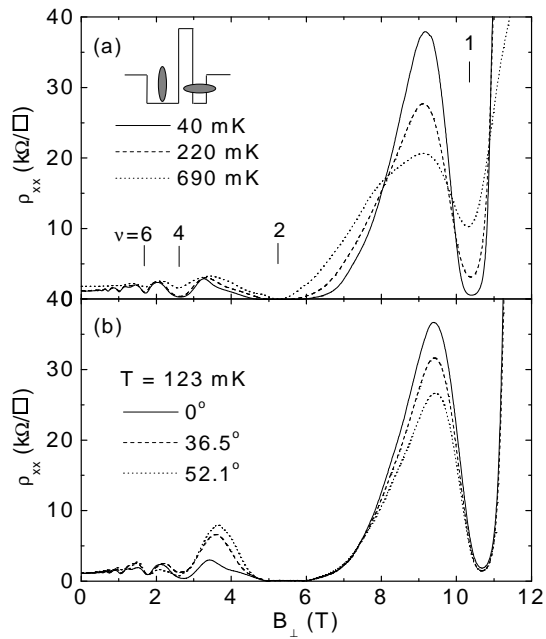}
    \caption{(a) Magneto-resistance, $\rho$$_{xx}$, versus perpendicular magnetic field for the balanced bilayer
    system, with {\it n}$_{N}$ = {\it n}$_{W}$ = 1.26x10$^{11}$ cm$^{-2}$, at three different temperatures, showing a $\nu$=1 QHS flanked by a reentrant insulating phase.
    (b) The dependence of $\rho$$_{xx}$ at balance on tilt angle.  Here, {\it n}$_{N}$ = {\it n}$_{W}$ = 1.30x10$^{11}$ cm$^{-2}$.}
\end{figure}

To examine the interlayer tunneling in our system, we measured
$\rho_{xx}$ traces in tilted magnetic fields [Fig.\ 3(b)].  The
component of the field applied parallel to the layers suppresses
tunneling by shifting the Fermi contours of the electrons in the
two layers with respect to one another in momentum space
\cite{hu92}. Consequently, one would expect that, as the parallel
field component is introduced, the strength of the $\nu$=1 minimum
will change if single-particle tunneling is playing a role in
stabilizing it. Even a small amount of tunneling can lead to
noticeable weakening of the $\nu$=1 QHS with parallel field
\cite{murphy94, yang94}.  The fact that the $\rho_{xx}$ minimum at
$\nu$=1 in our bilayer does not depend on the tilt angle and the
parallel field provides additional evidence that interlayer
tunneling in our system is small.

We conclude by making several remarks.  First, while all our data
point to the small value of tunneling in our bilayer system, we do
not know the strength of tunneling quantitatively. Theoretically,
the tunneling should be quite small; indeed, in an ideal system,
mixing between X$_{X}$ and X$_{Z}$ valleys is forbidden.  Optical
measurements in GaAs/AlAs superlattices, however, have indicated
that interface disorder can lead to some mixing \cite{tribe95}.
Direct measurements of tunneling in our system will therefore be
needed to assess the amount of tunneling quantitatively. Second,
we note that the difference in effective masses and well widths
for the layers implies that the ground state subband energies for
the two layers are different, even when the layers have equal
densities. This, together with the difference in g-factors, means
that the resulting Landau level fan diagram is rather complex. One
can achieve a situation, for example, where all the electrons in
one layer are spin-polarized while there is only partial
polarization in the other layer. Also, keeping the Fermi energies
of the layers equal as a function of sweeping magnetic field
requires some charge transfer between the wells. There are,
therefore, possibly interesting phenomena at high magnetic fields
that can be explored in this system \cite{zhu00}. Third, the
system has some remarkable properties even at zero or small
magnetic fields. For example, as shown in Fig.\ 2, one can
significantly tune the {\it g}*{\it m}* product in this system by
applying gate biases. A system with a gate-tunable {\it g}*{\it
m}* may find application in emerging fields such as quantum
computing \cite{divincenzo99}. Fourth, measurements of interlayer
Coulomb drag in a system where the electrons occupy different
valleys can be interesting. Finally, a bilayer system with
different effective masses in the two layers might host a novel
superconducting state \cite{macdonald03}.

We thank A.H. MacDonald, R. Winkler, and P.C. Klipstein for
helpful discussions, and the NSF and the Alexander von Humboldt
Foundation for financial support.  Part of the work was performed
at the Florida National High Magnetic Field Laboratory, which is
also supported by the NSF; we thank E. Palm, T. Murphy, and G.
Jones for technical support.

\break

\end{document}